\documentstyle[12pt]{article}

\begin{document}
\title{Prediction of three heavy spin-0 particles in the SM}
\author{Bing An Li\\
Department of Physics and Astronomy, University of Kentucky\\
Lexington, KY 40506, USA}

\maketitle
%\begin{center}
%{\bf hep-ph/0012051}
%\end{center}
\begin{abstract}
It is shown that in axial-vector field theory the axial-vector field
is always accompanied by a spin-0 field.
The perturbation theory of the axial-vector field theory is
reconstructed. These results are applied to the SM. One neutral and two
charged
spin-0 states are predicted, whose masses are \(m_{\phi^0}=m_t
e^{28.4}=3.78\times10^{14} GeV\) and \(m_{\phi^\pm}=m_t
e^{27}=9.31\times10^{13}GeV\) respectively.
A new perturbation theory of the SM is
proposed.
The propagators of Z and W fields in this new perturbation theory are
derived in unitary gauge. They have the same expressions as
the ones derived by using the renormalization gauge in the original perturbation
theory of the SM. No additional ghosts are associated with the new propagators.
The new propagators of W and Z fields show when energies are greater than
$10^{14}GeV$ the problems of indefinite metric and negative probability arise
in the SM.
\end{abstract}

\newpage
\section{Introduction}
The standard model[1] of electroweak interactions is
successful in many aspects.
The Lagrangian of the SM after spontaneous symmetry breaking is
\begin{eqnarray}
\lefteqn{{\cal L}=
-{1\over4}A^{i}_{\mu\nu}A^{i\mu\nu}-{1\over4}B_{\mu\nu}B^{\mu\nu}
+\bar{q}\{i\gamma\cdot\partial-m_q\}q}
\nonumber \\
&&+\bar{q}_{L}\{{g\over2}\tau_{i}
\gamma\cdot A^{i}+g'{Y\over2}\gamma\cdot B\}
q_{L}+\bar{q}_{R}g'{Y\over2}\gamma\cdot Bq_{R}\nonumber \\
&&+\bar{l}\{i\gamma\cdot\partial-m_{l}\}l
+\bar{l}_{L}\{{g\over2}
\tau_{i}\gamma\cdot A^{i}-{g'\over2}\gamma\cdot B\}
l_{L}-\bar{l}_{R}g'\gamma\cdot B l_{R}\nonumber \\
&&+{1\over2}m^2_ZZ_\mu Z^\mu+m^2_W W^+_\mu W^{-\mu}+{\cal L}_{Higgs}.
\end{eqnarray}
Summation over the fermion fields is
implicated in Eq.(1).

Because of parity nonconservation in weak interactions there
are both vector and axial-vector couplings between fermions and
gauge bosons. These couplings are written as
\begin{eqnarray}
\lefteqn{{\cal L}_{fi}={1\over4}\bar{q}\{g\tau_{i}
A^{i\mu}+g'YB^\mu\}\gamma_\mu (1+\gamma_5)
q+{g'\over4}\bar{q}Y\gamma_\mu (1-\gamma_5)qB^\mu}\nonumber \\
&&+{1\over4}\bar{l}\{g
\tau_{i}A^{i\mu}-g'B^\mu\}\gamma_\mu(1+\gamma_5)
l-{g'\over2}\bar{l}\gamma_\mu(1-\gamma_5)lB^\mu.
\end{eqnarray}

It is well known that in QED and QCD the
gauge bosons are pure vector fields.
However, the W and Z fields of the SM have both vector and axial-vector
components, as shown in Eq.(2).
In this paper the properties of the axial-
vector field is studied. New effects are found. Based on the findings
in the axial-vector field theory the SM is revisited. The paper
is organized as:1) introduction; 2) theory of axial-vector field;
3)necessity of revisiting SM;
4)neutral heavy spin-0 boson; 5)charged heavy spin-0 bosons; 6)
reconstruct the perturbation theory of the SM; 7)effect of vacuum
polarization by intermediate boson;
8)conclusion.

\section{Theory of axial-vector field}
In order to compare both vector and axial-vector field theories
the expression of the vacuum polarization
of a vector filed is presented first.
The Lagrangian of a vector field and a
fermion($QED$) is
\begin{equation}
{\cal L}=-{1\over4}(\partial_\mu v_\nu-\partial_\nu v_\mu)^2
+\bar{\psi}\{i\gamma\cdot
\partial+e\gamma\cdot v\}\psi-m\bar{\psi}\psi.
\end{equation}
This Lagrangian is invariant under the gauge transformation
\[\psi\rightarrow e^{i\alpha(x)}\psi,\;\;\; v_{\mu}\rightarrow
v_{\mu}+{1\over e}\partial_{\mu}\alpha.\]

The amplitude of the vacuum polarization is
\begin{eqnarray}
\lefteqn{<v|s|v>=i(2\pi)^4\delta(p'-p)\epsilon_\mu\epsilon_\nu\frac{e^2}
{(4\pi)^2}D
\Gamma(2-{D\over2})} \\
&&\int^1_0 dx
({\mu^2\over L_0})
^{{\epsilon\over2}}
\{x(1-x)(2p^\mu p^\nu-p^2 g^{\mu\nu})
+L_0 g^{\mu\nu}-m^2g^{\mu\nu}\}\nonumber \\
&&=
2i(2\pi)^4\delta(p'-p)\epsilon_\mu\epsilon_\nu\frac{e^2}
{(4\pi)^2}D\Gamma(2-{D\over2})(p^\mu p^\nu-p^2g^{\mu\nu})
\int^1_0 dx x(1-x)({\mu^2\over L_0})
^{{\epsilon\over2}},
\end{eqnarray}
where \(L_0=m^2-x(1-x)p^2\).
The amplitude is still gauge invariant.

The Lagrangian of a model of an axial-vector field and a
fermion is constructed as
\begin{equation}
{\cal L}=-{1\over4}(\partial_\mu a_\nu-\partial_\nu a_\mu)^2
+\bar{\psi}\{i\gamma\cdot
\partial+e\gamma\cdot a\gamma_{5}\}\psi-m\bar{\psi}\psi.
\end{equation}

The amplitude of the vacuum polarization is obtained
\begin{eqnarray}
\lefteqn{<a|s|a>=
i(2\pi)^4\delta(p'-p)\epsilon_\mu\epsilon_\nu\frac{e^2}{(4\pi)^2}D
\Gamma(2-{D\over2})}\nonumber \\
&&\int^1_0 dx({\mu^2\over L_0})
^{{\epsilon\over2}}\{x(1-x)(2p^\mu p^\nu-p^2 g^{\mu\nu})
+L_0 g^{\mu\nu}+m^2g^{\mu\nu}\} \\
&&=
2i(2\pi)^4\delta(p'-p)\epsilon_\mu\epsilon_\nu\frac{e^2}
{(4\pi)^2}D\Gamma(2-{D\over2})\nonumber \\
&&\int^1_0 dx \{x(1-x)(p^\mu p^\nu-p^2 g^{\mu\nu})+m^2 g^{\mu\nu}\}
({\mu^2\over L_0})
^{{\epsilon\over2}}.
\end{eqnarray}
In Eq.(7) the third term $m^2 g^{\mu\nu}$
has a plus sign, while in Eq.(4) this term has minus sign. The sign
difference is caused by the fact that $\gamma_5$ anticommutes
with $\gamma_\mu$. Because of this sign difference the mass term
is cancelled
in the case of vector field and Eq.(4) is obtained,
while for axial-vector field the expression(8) is obtained.
Comparing with Eq.(4),  in Eq.(8) there is one more term which is generated
dynamically.

Up to one loop the amplitude of the
vacuum polarization of vector field
is derived from Eq.(4)
\begin{equation}
\Pi^v_{\mu\nu}={1\over2}(p_\mu p_\nu-p^2 g_{\mu\nu})F_{v1}(z),
\end{equation}
where
\begin{eqnarray}
\lefteqn{F_{v1}(z)=1+
\frac{e^2}{(4\pi)^2}[{1\over3}D\Gamma(2-{D\over2})({\mu^2\over m^2})
^{\epsilon\over2}-8f_1(z)],}\\
&&f_1(z)=\int^1_0 dx x(1-x)log\{1-x(1-x)z\},
\end{eqnarray}
where \(z={p^2\over m^2}\).

The amplitude of the vacuum polarization of axial-vector field
is obtained from Eq.(8)
\begin{equation}
\Pi^a_{\mu\nu}={1\over2}(p_\mu p_\nu-p^2 g_{\mu\nu})F_{a1}(z)+F_{a2}(z)
p_\mu p_\nu+{1\over2}m^2_a
g_{\mu\nu},
\end{equation}
where
\begin{eqnarray}
\lefteqn{
F_{a1}(z)=1+
\frac{e^2}{(4\pi)^2}[{1\over3}D\Gamma(2-{D\over2})({\mu^2\over m^2})
^{\epsilon\over2}-8f_1(z)+8f_2(z)]\}\}} \\
&&f_2(z)={1\over z}\int^1_0 dx log\{1-x(1-x)z\}, \\
&&F_{a2}(z)=-\frac{4e^2}{(4\pi^2)}f_2(z), \\
&&m^2_a=
\frac{2e^2}{(4\pi^2)}D\Gamma(2-{D\over2})({\mu^2\over
m^2})^{\epsilon\over2}m^2.
\end{eqnarray}
Comparing with Eq.(9), there are two
new terms in Eq.(12): the mass term of the axial-vector field and
a term proportional
to $(\partial_\mu a^\mu)^2$. In Ref.[2] it has been already found that
mass of axial-vector field can be dynamically generated.
Both of the two new terms are generated by the vacuum polarization of the
axial-vector field.
The mechanism generating the new terms has been named as axial-vector
symmetry breaking
in Ref.[2].
Eq.(12) shows that vector and axial-vector field theories are very different.

The free and interaction Lagrangians of vector field theory(3) are defined as
\begin{eqnarray}
\lefteqn{{\cal L}_{v0}=
-{1\over4}(\partial_\mu v_\nu-\partial_\nu v_\mu)^2
+\bar{\psi}\{i\gamma\cdot
\partial-m\}\psi,}\\
&&{\cal L}_{vi}=e
\bar{\psi}\gamma_\mu
\psi v^\mu.
\end{eqnarray}
The massless vector field has two independent degrees of
freedom. After taking the vacuum polarization(4) into account, the degrees of
freedom of the vector field are still two. Therefore, the perturbation theory
(17,18)of the vector field theory satisfies unitarity.
As usual, the perturbation theory of axial-vector field
(6) is defined by
\begin{eqnarray}
\lefteqn{{\cal L}'_{a0}=
-{1\over4}(\partial_\mu a_\nu-\partial_\nu a_\mu)^2
+\bar{\psi}\{i\gamma\cdot
\partial-m\}\psi,}\\
&&{\cal L}'_{ai}=e
\bar{\psi}\gamma_\mu\gamma_5
\psi a^\mu.
\end{eqnarray}
Eq.(19) shows that
the axial-vector field is massless and \(\partial_\mu a^\mu=0.\). Therefore,
it has two independent degrees of freedom.
However,
after taking the vacuum polarization(8) into account, the axial-vector
field becomes massive and the term $F_{a2}p_\mu p_\nu$ in Eq.(12) indicates
that the divergence of
the axial-vector field is no longer zero. The vacuum polarization
makes the $a_\mu$ a field of four independent degrees of freedom.
Therefore,
the vacuum polarization of the axial-vector field cannot be treated
perturbatively and the perturbation theory of
axial-vector field theory must be reconstructed.

The function $F_{a1}$(13) is used to renormalize the $a_\mu$ field and
rewritten as(to $O(e^2)$)
\begin{eqnarray}
\lefteqn{F_{a1}(z)=Z_a\{1+(p^2-m^2_a)G_{a1}(p^2)\},}\\
&&Z_a=F_{a1}({m^2_a\over m^2}),\\
&&(p^2-m^2_a)G_{a1}(p^2)=F_{a1}(z)-F_{a1}({m^2_a\over m^2}),
\end{eqnarray}
where $Z_a$ is the renormalization constant of the field $a_\mu$,
$G_{a1}(p^2)$ is the radiative correction of the kinetic term.
Function
$F_{a2}$(15) is finite and rewritten as
\begin{equation}
F_{a2}(z)=\xi+(p^2-m^2_\phi)G_{a2}(p^2),
\end{equation}
where $m^2_\phi$ is the mass of a spin-0 state whose existence will be
studied below,
$G_{a2}$ is the radiative correction of the term $(\partial_\mu a^\mu)^2$, and
\begin{equation}
\xi=F_{a2}({m^2_\phi\over m^2}).
\end{equation}
The new perturbation theory of axial-vector field theory is constructed as
\begin{eqnarray}
\lefteqn{{\cal L}_{a0}=-{1\over4}(\partial_\mu a_\nu-\partial_\nu a_\mu)^2
+\xi(\partial_\mu a^\mu)^2+{1\over2}m^2_a a^2_\mu,}\\
&&{\cal L}_{ai}=e
\bar{\psi}\gamma_\mu\gamma_5\psi a^\mu+{\cal L}_c,\\
&&{\cal L}_c=
-\xi(\partial_\mu a^\mu)^2-{1\over2}m^2_a a^2_\mu.
\end{eqnarray}
${\cal L}_c$ is a counter term of the Lagrangian. Eq.(26) is the Stueckelberg's
Lagrangian.

The equation satisfied by $\partial_\mu a^\mu$ is derived from Eq.(26)
\begin{equation}
\partial^2(\partial_\mu a^\mu)-{m^2_a\over2\xi}(\partial_\mu a^\mu)=0.
\end{equation}
$\partial_\mu a^\mu$ is a pseudoscalar field and we define
\begin{equation}
\partial_\mu a^\mu=b\phi.
\end{equation}
The equation of the new field $\phi$ is found from Eq.(29)
\begin{equation}
\partial^2\phi-{m^2_a\over2\xi}\phi=0,\;\;\;m^2_{\phi}=-{m^2_a\over2\xi}.
\end{equation}
Eqs.(25,31) lead to that
the mass of the $\phi$ boson is the solution of the equation
\begin{equation}
2F_{a2}({m^2_\phi\over m^2})m^2_\phi+m^2_a=0.
\end{equation}
In order to show the existence of solution of Eq.(32)
we take
\[{e^2\over\pi^2}{m^2\over m^2_a}=1\]
as an example.
The numerical calculation shows that $\xi< 0$ and
the solution of Eq.(32) is found to be
\[m_\phi=8.02m.\]
If
\[{e^2\over\pi^2}{m^2\over m^2_a}=0.5\]
is taken we obtain
\[m_\phi=20.42 m.\]
The value of $m_\phi$ increases while $e^2$ decreases.
It is necessary to point out that $F_{a2}(z)$ is negative in the region of
the mass of $\phi$ boson.

We separate $a_\mu$ field into a massive spin-1 filed and a pseudoscalar
filed
\begin{eqnarray}
\lefteqn{a_\mu=a'_\mu+c\partial_\mu \phi,}\\
&&\partial_\mu a'^\mu=0.
\end{eqnarray}

Substituting Eq.(33) into Eq.(26),
the Lagrangian is divided into two parts
\begin{eqnarray}
\lefteqn{{\cal L}_{a0}={\cal L}_{a'0}+{\cal L}_{\phi 0},}\\
&&{\cal L}_{a'0}=-{1\over4}(\partial_\mu a'_\nu-\partial_\nu a'_\mu)^2
+{1\over2}m^2_a a'_\mu a'^\mu,\\
&&{\cal L}_{\phi 0}={1\over2m^2_\phi}\partial_\mu\phi\{
\partial^2+m^2_\phi\}\partial^\mu\phi.
\end{eqnarray}
The coefficient c is determined by the normalization of ${\cal L}_{\phi 0}$
\begin{equation}
c=\pm{1\over m_a},
\end{equation}
and substituting Eq.(33) into Eq.(30), we obtain
\begin{equation}
b=-cm^2_\phi=\mp{m^2_\phi\over m_a}.
\end{equation}
The sign of Eqs.(38,39) doesn't affect the physical
results when $\phi$ appears as virtual particle.
From Eq.(26) the propagator of the $a_\mu$ field is derived
\begin{equation}
\Delta_{\mu\nu}=
\frac{1}{p^2-m^2_a}\{-g_{\mu\nu}+(1+{1\over2\xi})
\frac{p_\mu p_\nu}{p^2-m^2_\phi}\}.
\end{equation}
Eq.(40) can be rewritten as
\begin{equation}
\Delta_{\mu\nu}=\frac{1}{p^2-m^2_a}\{-g_{\mu\nu}+\frac{p_\mu p_\nu}
{m^2_a}\}-{1\over m^2_a}\frac{p_\mu p_\nu}{p^2-m^2_\phi}.
\end{equation}
The first part is the propagator of the massive spin-1 field and
the second part is the propagator of the pseudoscalar field.
From Eqs.(33,41), the propagator of the pseudoscalar is determined to be
\begin{equation}
-\frac{1}{p^2-m^2_\phi}.
\end{equation}
It is different from the propagator of a regular spin-0 field by a minus sign.
The minus sign of Eq.(42) indicates
\begin{equation}
[a(k), a^\dag (k')]=-\delta_{kk'},
\end{equation}
where $a(k)$ and $a^\dag (k)$ are the annihilation and creation operators of
the
$\phi$ field respectively. The energy of the $\phi$ field is derived from
Eq.(37)
\begin{eqnarray}
\lefteqn{E=\int d^3 x{\cal H}=\sum_k \omega\{N_k+{1\over2}\},}\\
&&N_k=-a^\dag (k)a(k),
\end{eqnarray}
where $N_k$ is the number operator of the particle of momentum k
and \(\omega=\sqrt{k^2+m^2_\phi}\).
The energy(44) is positive.
In order to show that
Eq.(43) is understandable we take \(\xi=-{1\over2}\) in Eq.(26) as an example.
In the case of \(\xi=-{1\over2}\) both the spin-1 and spin-0 fields
have
the same mass.
Canonical quantization leads to
\begin{equation}
[a_\lambda (k), a^{\dag}_{\lambda'} (k')]=-g_{\lambda\lambda'}\delta_{kk'}.
\end{equation}
Eq.(46) is Lorentz covariant.
Taking \(\lambda, \lambda'=0\) the Eq.(43) is obtained.
Eq.(43) shows that there are
problems of indefinite metric and negative probability when the $\phi$ field
is on mass shell.

The coupling between the $\phi$ field and the fermion is found to be
\begin{equation}
\pm{e\over m_a}\bar{\psi}\gamma_\mu\gamma_5\psi\partial^\mu\phi=
\pm{e\over m_a}2im\bar{\psi}\gamma_5\psi\phi.
\end{equation}
The coupling is proportional to the fermion mass.

It is necessary to point out that the axial-vector field theory described by
Eq.(6) has triangle anomaly. As the SM we can add the second fermion with -e
as the coupling constant in Eq.(6), the triangle anomaly is cancelled out.
Because the number of the vertices in the vacuum polarization is even the
results obtained above still exist.

To summarize the results obtained in this section,
\begin{enumerate}
\item Two new terms, mass term of $a_\mu$ field and a term proportional to
$(\partial_\mu a^\mu)^2$,
are generated by the diagram of vacuum polarization of the
axial-vector field,
\item These two new terms make the $a_\mu$ a field of four independent
components: a massive spin-1 field and a pseudoscalar field,
\item The mass of the pseudoscalar is determined and the coupling between
$\phi$ field and fermion is proportional to the mass of the fermion,
\item The vacuum polarization cannot be
treated perturbatively. The new perturbation theory of the axial-vector field
(6) is reconstructed(26-28),
\item It is well known that the propagator of a
massive vector boson is expressed
as
\begin{equation}
\Delta^v_{\mu\nu}=\frac{1}{p^2-m^2_v}\{-g_{\mu\nu}+\frac{p_\mu p_\nu}{m^2_v}\}.
\end{equation}
This propagator causes quadratic divergence. However,
the propagator of massive
axial-vector boson(40) is very different from the propagator of massive vector
boson(48).
It doesn't cause quadratic divergence.
\item Eq.(41) shows that there is a pole at \(\sqrt{p^2}=m_{\phi}\). On the
other hand, the minus sign of Eq.(41) indicates when $\phi$ is on mass shell
there are problems of indefinite metric and negative probability in the
theory of axial-vector field.
\end{enumerate}
\section{Necessity of revisiting the SM}
As shown in eq.(2) both W and Z fields of the SM have axial-vector
components. Based on the results obtained in last section a revisit of the SM
is necessary. The issue is how to do perturbation in the SM after spontaneous
symmetry breaking.
Using the unitary gauge, after spontaneous symmetry breaking
the free Lagrangian of W and Z bosons
in the original perturbation theory is defined as
\begin{equation}
{\cal L}_0=-{1\over4}(\partial_\mu Z\nu-\partial_\nu Z_\mu)^2+{1\over2}
m^2_Z Z_\mu Z^\mu-{1\over2}(\partial_\mu W^+_\nu-\partial_\nu W^+_\mu)
(\partial^\mu W^{-\nu}-\partial^\nu W^{-\mu})+m^2_W W^+_\mu W^{-\mu}.
\end{equation}
There are others
\[{\cal L}_{\gamma0}+{\cal L}_{f0}+{\cal L}_{H0},\]
where ${\cal L}_{\gamma0}$, ${\cal L}_{f0}$ and ${\cal L}_{H0}$ are
the free Lagrangians
of photon, fermions and Higgs respectively.
The interaction Lagrangian can be found from Eq.(1).
If renormalization gauge is chosen there is gauge fixing term in Eq.(49).
However, ghost is associated.
In Eq.(49) both W and Z are massive spin-1
fields.

According to the study of section(2) the axial-vector couplings between
the intermediate bosons and fermions(2) lead to the existence of
one neutral
and two charged spin-0 states, which are associated with Z and
$W^\pm$ respectively.
In the free Lagrangian of the intermediate bosons(49)
there are no such spin-0
state. Therefore,
the free Lagrangian of
intermediate bosons should be redefined and the
perturbation
theory of the SM should be reconstructed. Therefore, revisit of the SM is
necessary.

In the SM the intermediate bosons are Yang-Mills fields.
Besides the vacuum polarization by fermions the intermediate bosons
contribute to the vacuum polarization too. However, the calculation of the
contribution of the intermediate bosons to the vacuum polarization can only
be done after the propagators of intermediate bosons are defined by
the new free Lagrangian of the intermediate bosons in the new perturbation
theory. In section(7) the issue will be addressed.

\section{Neutral heavy spin-0 state}
The $F_{1,2}$ functions defined in section(2) can be found from
the vacuum polarization of Z boson by fermions.

We start from the t- and b-quark generation. From the SM(1) the Lagrangian
of the interactions between Z-boson and t and b quarks is found
\begin{equation}
{\cal L}={\bar{g}\over4}\{
(1-{8\over3}\alpha)\bar{t}\gamma_\mu t+
\bar{t}\gamma_\mu\gamma_5  t\}Z^\mu
-{\bar{g}\over4}\{(1-{4\over3}\alpha)\bar{b}\gamma_\mu b
+\bar{b}\gamma_\mu\gamma_5 b\}Z^\mu,
\end{equation}
where \(\alpha=sin^2 \theta_W \).
The S-matrix element of the vacuum polarization of t and b quark generation
at the second order
is obtained
\begin{eqnarray}
\lefteqn{<Z|s^{(2)}|Z>=i(2\pi)^4\delta(p-p')\epsilon^\mu
\epsilon^\nu
{\bar{g}^2\over8}
\frac{N_C}{(4\pi)^2}D\Gamma(2-{D\over2})}\nonumber\\
&&\int^1_0 dx\{
x(1-x)(p_\mu p_\nu-p^2 g_{\mu\nu})[
({\mu^2\over L_t})^{{\epsilon\over2}}	
[(1-{8\over3}\alpha)^2+1]+
({\mu^2\over L_b})^{{\epsilon\over2}}	
[(1-{4\over3}\alpha)^2+1]]\nonumber \\
&&+m^2_t
({\mu^2\over L_t})^{{\epsilon\over2}}g_{\mu\nu}+m^2_b	
({\mu^2\over L_b})^{{\epsilon\over2}}g_{\mu\nu}\},
\end{eqnarray}
where
\(L_t=m^2_t-x(1-x)p^2\) and	
\(L_b=m^2_b-x(1-x)p^2\).	
The S-matrix elements of other two generation of quarks are
obtained too. The kinetic term of Eq.(51) is generated by both the vector and
the axial-vector couplings and the mass terms originate in the axial-vector
coupling only.

The interaction Lagrangian between Z-boson and the leptons of e and $\nu_e$ is
obtained from
Eq.(1)
\begin{equation}
{\cal L}={\bar{g}\over4}
\bar{\nu_e}\gamma_\mu(1+\gamma_5) \nu_e
Z^\mu
-{\bar{g}\over4}\{(1-4\alpha)\bar{e}\gamma_\mu e
+\bar{e}\gamma_\mu\gamma_5 e\}Z^\mu.
\end{equation}
We obtain
\begin{eqnarray}
\lefteqn{<Z|s^{(2)}|Z>=i(2\pi)^4\delta(p-p'){\bar{g}^2\over8}\epsilon^\mu
\epsilon^\nu\frac{1}{(4\pi)^2}D\Gamma(2-{D\over2})}\nonumber \\
&&\int^1_0 dx\{
x(1-x)(p_\mu p_\nu-p^2 g_{\mu\nu})[
({\mu^2\over L_e})^{{\epsilon\over2}}	
[(1-4\alpha)^2+1]+
2({\mu^2\over L_\nu})^{{\epsilon\over2}}]\nonumber	\\
&&+m^2_e
({\mu^2\over L_e})^{{\epsilon\over2}}g_{\mu\nu}+m^2_\nu	
({\mu^2\over L_\nu})^{{\epsilon\over2}}g_{\mu\nu}\},
\end{eqnarray}
There are other two lepton generations contributing to the vacuum
polarization.

The amplitude of the vacuum polarization of fermions is expressed as
\begin{equation}
\Pi^Z_{\mu\nu}={1\over2}F_{Z1}(z)(p_\mu p_\nu-p^2 g_{\mu\nu})+F_{Z2}(z)
p_\mu p_\nu+{1\over2}\Delta m^2_Z g_{\mu\nu},
\end{equation}
\begin{eqnarray}
\lefteqn{F_{Z1}=1+\frac{\bar{g}^2}{64\pi^2}\{\frac{D}{12}\Gamma(2-{D\over2})
[N_C y_q\sum_q(\frac{\mu^2}{m^2_q})^{{\epsilon\over2}}
+y_l\sum_l(\frac{\mu^2}{m^2_l})^{{\epsilon\over2}}]}\nonumber \\
&&-2[N_C y_q\sum_q f_1(z_q)+y_l\sum_lf_1(z_l)]+2[\sum_q f_2(z_q)
+\sum_{l=e,\mu,\tau}
f_2(z_l)]\},
\nonumber \\
&&F_{Z2}=-\frac{\bar{g}^2}{64\pi^2}\{N_C\sum_qf_2(z_q)
+\sum_{l=e,\mu,\tau}f_2(z_l)\},\\
&&\Delta m^2_Z={1\over8}{\bar{g}^2\over(4\pi)^2}D\Gamma(2-{D\over2})
\{N_c\sum_{q}m^2_q
({\mu^2\over m^2_q})^{{\epsilon\over2}}+\sum_l m^2_l({\mu^2\over m^2_l})^
{{\epsilon\over2}}\}\nonumber .
\end{eqnarray}
where \(y_q=1+(1-{8\over3}\alpha)^2\) for \(q=t,c,u\),
\(y_q=1+(1-{4\over3}\alpha)^2\)
for
\(q=b,s,d\), \(y_l=1+(1-4\alpha)^2\), for \(l=\tau, \mu, e\), \(y_l=2\) for
\(l=\nu_e,\nu_\mu,\tau_\mu\),
\(z_i={p^2\over m^2_i}\) and the functions $f_{1,2}$
are defined by Eqs.(11,14) respectively.
In the SM the Z boson gains mass from the spontaneous symmetry breaking
and the mass term $\Delta m^2_Z$
has been refereed to the renormalization of $m^2_Z$. In Ref.
2Ythe
mechanism of
generating $m_Z$ and $m_W$ is explored.
Both the vector and axial-vector couplings of Eq.(2) contribute to $F_{Z1}$.
Only the axial-vector coupling of Eq.(2) contributes to $F_{Z2}$. $F_{Z2}$
is finite.

The function $F_{Z1}$ is used to renormalize the Z-field. We are
interested in $F_{Z2}$ and it
is rewritten as
\begin{equation}
F_{Z2}(z)=\xi_Z+(p^2-m^2_{\phi^0})G_{Z2}(p^2),
\end{equation}
where $m^2_{\phi^0}$ is the mass of a new neutral spin-0 field, $\phi^0$,
which will
be studied, $G_{Z2}$ is the radiative correction of this term, and
\begin{equation}
\xi_Z
=F_{Z2}|_{p^2=m^2_{\phi^0}}.
\end{equation}
Now we reconstruct the perturbation theory. The free Lagrangian of the Z-field
is defined as
\begin{equation}
{\cal L}_{Z0}=-{1\over4}(\partial_\mu Z_\nu-\partial_\nu
Z_\mu)^2+\xi_Z(\partial_\mu Z^\mu)^2+{1\over2}m^2_Z Z^2_\mu.
\end{equation}	
The equation of $\partial_\mu Z^\mu$ is derived from Eq.(58)
\begin{equation}
\partial^2(\partial_\mu Z^\mu)-{m^2_Z\over2\xi_Z}(\partial_\mu Z^\mu)=0.
\end{equation}
$\partial_\mu Z^\mu$ is an independent spin-0 field.
Following section(2) we have
\begin{eqnarray}
\lefteqn{Z_\mu=Z'_\mu\pm{1\over m_Z}\partial_\mu\phi^{0},}\\
&&\partial_\mu Z'^\mu=0,\\
&&\phi^0=\mp{m_Z\over m^2_{\phi^0}}\partial_\mu Z^\mu,\\
&&\partial^2\phi^0-{m^2_Z\over2\xi_Z}\phi^0=0.
\end{eqnarray}
The mass of $\phi^0$ is obtained
\begin{eqnarray}
\lefteqn{2m^2_{\phi^0}F_{Z2}|_{p^2=m^2_{\phi^0}}+m^2_Z=0,}\\
&&m^2_{\phi^0}=-{m^2_Z\over2\xi_{Z}}.
\end{eqnarray}
Using Eq.(55), Eq.(64) is rewritten as
\begin{equation}
3\sum_q{m^2_q\over m^2_Z}z_q f_2(z_q)+\sum_l {m^2_l\over m^2_Z}z_l
f_2(z_l)={32\pi^2\over\bar{g}^2}.
\end{equation}
For $z>4$ it is found from Eq.(14) that
\begin{equation}
f_2(z)=-{2\over z}-{1\over z}
(1-{4\over z})^{{1\over2}}log\frac{1-(1-{4\over z})^{{1\over2}}}
{1+(1-{4\over z})^{{1\over2}}}.
\end{equation}
Because of the ratios of ${m^2_q\over m^2_Z}$ and ${m^2_l\over m^2_Z}$
top quark dominates the Eq.(66) and the contributions of other fermions
can be
ignored. Eq.(66) has a solution at very large value of z. For very large z
Eq.(66)
becomes
\begin{equation}
\frac{2(4\pi)^2}{\bar{g}^2}+\frac{6m^2_t}{m^2_Z}=3{m^2_t\over m^2_Z}log
{m^2_{\phi^0}\over m^2_t}.
\end{equation}
The mass of the $\phi^0$ is determined to be
\begin{equation}
m_{\phi^0}=m_t e^{\frac{m^2_z}{m^2_t}{16\pi^2\over3\bar{g}^2}+1}
=m_t e^{28.4}=3.78\times10^{14}GeV,
\end{equation}
and
\[\xi_Z=-1.18\times10^{-25}.\]
The neutral spin-0 boson is extremely heavy.

The propagator of Z boson is found from Eq.(58)
\begin{equation}
\Delta_{\mu\nu}=
\frac{1}{p^2-m^2_Z}\{-g_{\mu\nu}+(1+\frac{1}{2\xi_Z})\frac{p_\mu p_\nu}{
p^2-m^2_{\phi^0}}\},
\end{equation}
It can be separated into two parts
\begin{equation}
\Delta_{\mu\nu}=
\frac{1}{p^2-m^2_Z}\{-g_{\mu\nu}+\frac{p_\mu p_\nu}{
m^2_Z}\}-\frac{1}{m^2_Z}\frac{p_\mu p_\nu}{p^2
-m^2_{\phi^0}}.
\end{equation}
The first part is the propagator of the physical spin-1 Z boson and
the second part is the propagator of a new
neutral spin-0 meson, $\phi^0$.

It is well known that the propagator of Z boson of the original perturbation
theory
takes the same form as Eq.(70)
in the renormalization gauge. However,
from
physical point of view they are different.
The SM is gauge invariant before spontaneous symmetry breaking. Therefore,
in general, there is a gauge fixing term in the Lagrangian of the SM. In the
study presented above the gauge parameter has been chosen to be zero, unitary
gauge.
The differences between Eq.(70) and the propagator of renormalization gauge
in the original perturbation theory of the SM are
\begin{enumerate}
\item In eq.(70) the $\xi_z$ is determined by Eqs.(57,69) dynamically.
Physics results depend on it.
The gauge parameter of the renormalization gauge is determined by choosing
gauge and because of unitarity
physics results are independent of the gauge parameter.
\item In renormalization gauge ghosts are accompanied. However,
there are no additional ghosts associated with
Eqs.(70,71).
\end{enumerate}

The propagator(70) shows that when $p^2<<m^2_{\phi^0}$ it takes
\begin{equation}
\Delta_{\mu\nu}=
\frac{1}{p^2-m^2_Z}\{-g_{\mu\nu}+\frac{p_\mu p_\nu}{m^2_Z}\}.
\end{equation}
Because of the huge value of $m^2_{\phi^0}$ Eq.(72) is always a good
approximation. Only for extremely high momentum the complete propagator(70)
is needed.

Eq.(71) shows that there is a pole at \(p^2=m^2_{\phi^0}\). On the other hand,
the minus sign of Eq.(71) indicates that the
annihilation and creation operators of free $\phi^0$ field obey Eq.(43).
Therefore, the Fock space has indefinite metric and there is problem of
negative probability when $\phi^0$ is on mass shell.

Substituting Eq.(60) into Eqs.(50,52), the couplings between
$\partial_\mu \phi^0$
and
t,b,e,$\nu_e$ fermions are found
\begin{eqnarray}
\lefteqn{{\cal L}=\pm{1\over m_Z}{\bar{g}\over4}\{
(1-{8\over3}\alpha)\bar{t}\gamma_\mu t+
\bar{t}\gamma_\mu\gamma_5  t\}\partial_\mu\phi^0}\nonumber \\
&&\pm{1\over m_Z}{\bar{g}\over4}\{-(1-{4\over3}\alpha)\bar{b}\gamma_\mu b
-\bar{b}\gamma_\mu\gamma_5 b\}\partial_\mu\phi^0\nonumber \\
&&\pm{1\over m_Z}{\bar{g}\over4}\{
\bar{\nu_e}\gamma_\mu(1+\gamma_5) \nu_e
-(1-4\alpha)\bar{e}\gamma_\mu e
-\bar{e}\gamma_\mu\gamma_5 e\}\partial_\mu \phi^0.
\end{eqnarray}
The couplings with other generations of fermions take the same form.
Using the equations of fermions, it can be found that
the couplings between $\phi^0$ and fermions are
\begin{equation}
\pm{1\over m_Z}{\bar{g}\over4}2i\sum_i m_i\bar{\psi}_i\gamma_5\psi_i \phi^0,
\end{equation}
where i stands for the type of fermion.
Eq.(74) shows that the coupling between
$\phi^0$ and fermion is proportional to the mass of the fermion. There are
other couplings obtained from Eq.(73). The couplings between $\phi^0$ and
intermediate bosons can be obtained too.

\section{Two heavy charged spin-0 fields}
In the SM the fermion-W vertices are
\begin{equation}
{\cal L}={g\over4}\bar{\psi}\gamma_\mu(1+\gamma_5)\tau^i\psi W^{i\mu},
\end{equation}
where $\psi$ is the doublet of fermion and summation over all fermion
generations is
implicated.

Using the vertices(75),
the expression of the vacuum polarization of fermions is obtained
\begin{equation}
\Pi^W_{\mu\nu}=F_{W1}(p^2)(p_\mu p_\nu-p^2 g_{\mu\nu})+2F_{W2}(p^2)
p_\mu p_\nu+\Delta m^2_W
g_{\mu\nu},
\end{equation}
where
\begin{eqnarray}
\lefteqn{F_{W1}(p^2)=1+{g^2\over32\pi^2}D\Gamma(2-{D\over2})\int^1_0 dx
x(1-x)\{
N_C\sum_{iq}({\mu^2\over L^i_q})^{{\epsilon\over2}}+
\sum_{il}({\mu^2\over L^i_l})^{{\epsilon\over2}}\}}\nonumber \\
&&-{g^2\over16\pi^2}\{N_C\sum_{iq} f^i_{1q}+\sum_{il}
f^i_{1l}\}+{g^2\over16\pi^2}\{N_C\sum_{iq} f^i_{2q}+\sum_{il}f_{2l}\},\\
&&F_{W2}(p^2)=-{g^2\over32\pi^2}\{N_C\sum_{iq} f^i_{2q}+\sum_{il}
f^i_{2l}\},\\
&&
\Delta m^2_W={g^2\over4}{1\over(4\pi)^2}D\Gamma(2-{D\over2})
\int^1_0 dx\{N_c\sum_{iq}L^i_q
({\mu^2\over L^i_q})^{{\epsilon\over2}}+\sum_{il}L^i_l({\mu^2\over L^i_l})^
{{\epsilon\over2}}\}.
\end{eqnarray}
where
\begin{equation}
L^1_q =m^2_b x+m^2_t (1-x),\;\;
L^2_q =m^2_s x+m^2_c (1-x),\;\;
L^3_q =m^2_d x+m^2_u (1-x),
\end{equation}
\[L^1_l =m^2_e x,\;\;
L^2_l =m^2_\mu x,\;\;
L^3_l =m^2_\tau x,\]
\begin{eqnarray}
\lefteqn{
f^i_{1q}=\int^1_0 dx x(1-x)log[1-x(1-x){p^2\over L^i_q}]},\\
&&f^i_{1l}=\int^1_0 dx x(1-x)log[1-x(1-x){p^2\over L^i_l}],\\
&&f^i_{2q}={1\over p^2}\int^1_0 dx L^i_q log[1-x(1-x)
{p^2\over L^i_q}],\\
&&f^i_{2l}={1\over p^2}\int^1_0 dx L^i_l log[1-x(1-x){p^2\over L^i_l}].
\end{eqnarray}
Function $F_{W1}(p^2)$ is used to renormalize the W-field.
In the SM W boson gains mass from spontaneous symmetry breaking and the
additional mass term(79) has been treated by renormalization.
Function $F_{W2}$ leads to the existence of two charged spin-0 states,
$\phi^\pm$,
in the SM. $F_{W2}$ is rewritten as
\begin{eqnarray}
\lefteqn{F_{W2}=\xi_W+(p^2-m^2_{\phi_W})G_{W2}(p^2),}\\
&&\xi_W=F_{W2}(p^2)|_{p^2=m^2_{\phi_W}},
\end{eqnarray}
where $G_{W2}$ is the radiative correction of the term
$(\partial_\mu W^\mu)^2$ and $m^2_{\phi_W}$ is the mass of the charged spin-0
states, $\phi^\pm$, whose existence will be shown below.

The free part of the Lagrangian of W-field
is redefined as
\begin{equation}
{\cal L}_{W0}=-{1\over2}(\partial_\mu W^+\nu-\partial_\nu W^+_\mu)
(\partial_\mu W^-_\nu-\partial_\nu W^-_\mu)
+2\xi_W\partial_\mu W^{+^\mu}
\partial_\nu W^{-^\nu}
+m^2_W W^+_\mu W^{-\mu}.
\end{equation}	
From Eq.(87) the equation satisfied by the divergence of the W-field
is derived
\begin{equation}
\partial^2(\partial_\mu W^{\pm\mu})-{m^2_W\over2\xi_W}(\partial_\mu
W^{\pm\mu})=0.
\end{equation}
$\partial_\mu W^{\pm\mu}$ are spin-0 fields. Therefore,
the W field of the SM has four independent
components.
The W-field is decomposed as
\begin{eqnarray}
\lefteqn{W^\pm_\mu=W'^\pm_\mu \pm{1\over m_W}\partial_\mu \phi^\pm,}\\
&&\partial\mu W'^{\pm\mu}=0,\\
&&\phi^\pm=\mp{m_W\over m^2_{\phi_W}}\partial_\mu W^{\pm\mu}.
\end{eqnarray}
The equation of $\phi^\pm$ is derived from Eqs.(88,91)
\begin{equation}
\partial^2\phi^\pm-{m^2_W\over2\xi_{W}}\phi^\pm=0.
\end{equation}

It is the same as Eq.(64) the
mass of $\phi^\pm$ is determined by the equation
\begin{equation}
2m^2_{\phi_W}F_{W2}(p^2)|_{p^2=m^2_{\phi_W}}+m^2_W=0
\end{equation}
and
\begin{equation}
m^2_{\phi_W}=-{m^2_W\over 2\xi_{W}}.
\end{equation}
Numerical calculation shows that top quark is dominant in $F_{W2}$. Keeping the
contribution of top quark only, Eq.(78) becomes
\begin{equation}
{p^2\over m^2_W}F_{W2}=-\frac{3g^2}{32\pi^2}{m^2_t\over m^2_W}\{-{3\over4}
+{1\over2z}
+[{1\over2}-{1\over z}+{1\over2z^2}]log(z-1)\},
\end{equation}
where \(z={p^2\over m^2_t}\).
Eq.(93) has a solution at very large z. At very large z
Eq.(95) becomes
\begin{equation}
{p^2\over m^2_W}F_{W2}=-{3g^2\over64\pi^2}{m^2_t\over m^2_W}logz.
\end{equation}
The mass of $\phi^\pm$ is determined to be
\begin{equation}
m_{\phi_W}=m_t e^{{16\pi^2\over3g^2}{m^2_W\over m^2_t}}=m_t e^{27}
=9.31\times10^{13}GeV,
\end{equation}
and
\[\xi_W=-3.73\times10^{-25}.\]
The charged $\phi^\pm$ are very heavy too.

The propagator of W-field is derived from Eq.(87)
\begin{equation}
\Delta^W_{\mu\nu}=
\frac{1}{p^2-m^2_W}\{-g_{\mu\nu}+(1+\frac{1}{2\xi_W})\frac{p_\mu p_\nu}{
p^2-m^2_{\phi_W}}\},
\end{equation}
and it can be separated into two parts
\begin{equation}
\Delta^W_{\mu\nu}=
\frac{1}{p^2-m^2_W}\{-g_{\mu\nu}+\frac{p_\mu p_\nu}{
m^2_W}\}-\frac{1}{m^2_W}\frac{p_\mu p_\nu}{p^2
-m^2_{\phi_W}}.
\end{equation}
The first part of Eq.(99) is the propagator of physical
spin-1 W-field and the second
part is related to the propagator of the $\phi^\pm$ field. The same as
mentioned in section(5) the physics of the propagator(98) is different from
the one derived by using
renormalization gauge in the original perturbation theory
of the SM. When $p^2<<m^2_{\phi^\pm}$ we have
\begin{equation}
\Delta^W_{\mu\nu}=\frac{1}{p^2-m^2_W}(-g_{\mu\nu}+\frac{p_\mu p_\nu}{m^2_W}).
\end{equation}
In most cases Eq.(100) is a very good approximation.

Eq.(99) shows that at very high energy there is a pole at
\(\sqrt{p^2}=m_{\phi_W}\). On the other hand, the minus sign in Eq.(99)
indicates that there are problems of indefinite metric and negative
probability when $\phi^{\pm}$ are on mass shell.

The Lagrangian of interactions between fermions and $\partial_\mu \phi^\pm$
field is
found from Eqs.(75,89)
\begin{equation}
{\cal L}_{q\phi}=\pm{1\over m_W}{g\over4}\sum_j\bar{\psi}_j
\gamma_\mu(1+\gamma_5)
\tau^i\psi_j
\partial^\mu\phi^{i},
\end{equation}
where j is the type of the fermion and
\begin{equation}
\phi^1={1\over\sqrt{2}}(\phi^+ +\phi^-)\;\;\;
\phi^2={1\over \sqrt{2}i}(\phi^+ -\phi^-).
\end{equation}
Using the dynamical equations of fermions of the SM, for t and b quark
generation we obtain
\begin{equation}
{\cal L}'_{q\phi}=\pm{i\over m_W}{g\over4}(m_t+m_b)
\{\bar{\psi}_t\gamma_5\psi_b\phi^+
+\bar{\psi}_b\gamma_5\psi_t\phi^-\}.
\end{equation}
The coupling is proportional to the fermion mass. The couplings between other
generations of
fermion and $\phi^\pm$ field are the same as Eq.(103).
\section{Reconstruct the perturbation theory of the SM}
We have learned from sections(4,5) that when the
vacuum polarization of W and Z
fields by ferminos
are treated nonperturbatively three heavy spin-0 states,$\phi^0$
and $\phi^\pm$,
 are revealed. Therefore, the W and Z of the SM(1) are the fields of four
independent components. They are composed of spin-1 and spin-0 states.
In the original perturbation theory of the SM there are no such
spin-0 states. Therefore,
the perturbation theory of the SM must be reconstructed.
The changes are made in the free Lagrangian of W
and Z fields and related counter terms. All other parts of the Lagrangian
are kept the same as in the original perturbation theory of the SM.
The new perturbation theory of the SW is constructed as
\begin{enumerate}
\item
Based on the study of sections(4,5) the new free Lagrangian of the W and Z
fields
is constructed as
\begin{eqnarray}
\lefteqn{{\cal L}_0=-{1\over4}(\partial_\mu Z\nu-\partial_\nu Z_\mu)^2
+\xi_Z(\partial_\mu Z^\mu)^2+{1\over2}m^2_Z Z_\mu Z^\mu}\nonumber \\
&&-{1\over2}
(\partial_\mu W^+_\nu-\partial_\nu W^+_\mu)
(\partial_\mu W^-_\nu-\partial_\nu W^-_\mu)+2\xi_W\partial_\mu W^{+\mu}
\partial_\nu W^{-\nu}
+m^2_W W^+_\mu W^{-\mu}.
\end{eqnarray}
\item
Counter terms must be added to the interaction part of the Lagrangian of
the SM
\begin{equation}
{\cal L}_c=-\xi_Z(\partial_\mu Z^\mu)^2-2\xi_W\partial_\mu W^{+\mu}\partial_\nu
W^{-\nu}.
\end{equation}
\item The propagators of Z and W fields are expressed as Eqs.(70,98). There are
no ghosts which associate with Eqs.(70,98).
When momentum is much
less than the mass of the spin-0 state the propagators(70,98) go back to
Eqs.(72,100). Because of the values of $m^2_{\phi^0}$ and $m^2_{\phi_W}$
are so large. Eqs.(72,100) are good approximations. However, for loop diagrams
in which propagators of intermediate bosons are involved the propagators(70,98)
must be taken.
\item
The existence of three heavy spin-0 states are predicted.
The masses of
$\phi^0$ and $\phi^\pm$ are determined by Eqs.(69,97). The interactions
between $\phi^0$ and $\phi^\pm$ and fermions are proportional to the
masses of
fermions(74,103). The interactions of $\phi^0$ and $\phi^\pm$ with
Z' and W' can be obtained too.
\item The free Lagrangian(104) is in the unitary gauge.
Renormalization gauge can be used too. However, the expression of
the propagators(70,98) show that the choice of renormalization gauge is
not necessary.
\end{enumerate}
\section{Effects of vacuum polarization by intermediate bosons}
Besides the vacuum polarization by fermions
the intermediate bosons contribute to the vacuum polarization too.
Before we proceed to study the effects of the vacuum polarization by
intermediate bosons it is necessary to restate the theoretical approach
exploited in this paper. In the SM there are two kinds of
vacuum polarization of W and Z: vacuum polarization by fermions and by
intermediate bosons themselves.
As shown in sections(4,5) the
free Lagrangian of the intermediate bosons needs to be reconstructed.
At the lowest order the propagators of fermions are not affected and the
vacuum polarization of W and Z by fermions are calculated in sections(4,5).
However, in the new perturbation theory
the vacuum polarization by the intermediate bosons can be
calculated only after the propagators of W and Z are defined.

After the propagators of Z and W are defined(70,98) we can proceed to
study the effects of the vacuum polarization by intermediate bosons.
The interaction Lagrangian of intermediate bosons is obtained form the
SM
\begin{eqnarray}
\lefteqn{{\cal L}_i=i\bar{g}(\partial_\mu Z_\nu-\partial_\nu Z_\mu)
W^{-\mu}W^{+\nu}}\nonumber \\
&&+i\bar{g}Z^\mu\{(\partial_\mu W^+_\nu-\partial_\nu W^+_\mu)W^-_\nu
-(\partial_\mu W^-_\nu-\partial_\nu W^-_\mu)W^+_\nu\}\nonumber \\
&&-\bar{g}^2\{Z_\mu Z^\mu W^+_\nu W^{-\nu}-Z_\mu Z_\nu W^{+\mu}W^{-\nu}\}.
\end{eqnarray}
We calculate the contribution of W bosons to the vacuum polarization of
Z boson.
The same as Eq.(54)
in the amplitude of the vacuum polarization of Z boson there are three
parts: kinetic term, mass term, and the term proportional to $p_\mu p_\nu$,
$F'_{Z2}p_\mu p_\nu$.
We
are interested in the last term. The
calculation shows that only the second term of the Lagrangian(106)
contributes to $F'_{Z2}$. Using the propagator of W boson(98) and Eq.(106),
we obtain
\begin{eqnarray}
\lefteqn{F'_{Z2}=\frac{2\bar{g}^2}{(4\pi)^2}\{{1\over4}\Gamma(2-{D\over2})
({\mu^2\over m^2_W})^{{\epsilon\over2}}-{1\over12}+{3\over2}
\int^1_0 dx[{3\over z}x(1-x)-x(5-7x)]log[1-x(1-x)z]\}}\nonumber \\
&&+\frac{2\bar{g}^2 b}{(4\pi)^2}\{-{13\over40}+{3\over2}\int^1_0 dx x^3 (1-x)
[x(m^2_{\phi_W}-m^2_W)+m^2_W][x(m^2_{\phi_W}-m^2_W)+m^2_W-x(1-x)p^2]^{-1}
\nonumber \\
&&-{3\over bm^2_W}\int^1_0 dxx(2x-{3\over2})[m^2_W-x(1-x)p^2]
log\{[m^2_W-x(1-x)p^2]\nonumber \\
&&[x(m^2_{\phi_W}-m^2_W)+m^2_W-x(1-x)p^2]^{-1}\}
\nonumber \\
&&-{9\over2}{1\over p^2}\int^1_0 dx\int^x_0 dy[y(m^2_{\phi_W}-m^2_W)+m^2_W]
log\{[y(m^2_{\phi_W}-m^2_W)+m^2_W-x(1-x)p^2]\nonumber \\
&&[y(m^2_{\phi_W}-m^2_W)+m^2_W]^{-1}\},
\end{eqnarray}
where \(z={p^2\over m^2_W}\) and \(b=1+{1\over2\xi_W}\).

There is divergence in Eq.(107), therefore, renormalization of the operator
$(\partial_\mu Z^\mu)^2$ is required. As Eq.(56) $F'_{Z2}$ is written as
\begin{equation}
F'_{Z2}=F'_{Z2}(m^2_{\phi^0})+(p^2-m^2_{\phi^0})G'_{Z2}(p^2).
\end{equation}
$F'_{Z2}$ is divergent and
$G'_{Z2}$ is finite and another term of radiative correction
of $(\partial_\mu Z^\mu)^2$. We
define
\begin{eqnarray}
\lefteqn{\xi_Z+F'_{Z2}(m^2_{\phi^0})=\xi_Z Z_Z,}\\
&&Z_Z=1+{1\over \xi_Z}F'_{Z2}(m^2_{\phi^0}).
\end{eqnarray}
$Z_Z$ is the renormalization constant of the operator
$(\partial_\mu Z^\mu)^2$.
The renormalization constant $Z_Z$ defined in Eq.(110)
guarantees the Z boson
the same propagator(70).

In the same way, the contribution of Z and W bosons to the vacuum
polarization of W
boson can be calculated and the renormalization constant $Z_W$ can be
defined.
After renormalization of $\partial_\mu W^{+\mu}\partial_\nu W^{-\nu}$
we still have
the same propagator of W boson(98).
\section{Conclusion}
To summarize the results obtained in this paper,
\begin{enumerate}
\item After taking vacuum polarization of one loop into account,
a spin-0 state
is revealed from a axial-vector field theory.
The
perturbation theory of axial-vector field theory is reconstructed.
\item The W and Z fields of the SM are different from photon and gluon
fields.
Besides vector components they have axial-vector components.
After treating the
vacuum polarization of W and Z by fermions nonperturbatively,
the W and Z of
the SM are composed of spin-1 and spin-0 fields.
\item One neutral and two charged spin-0 states are revealed from the SM.
They
are very heavy, \(m_{\phi^0}=3.78\times10^{14}GeV\),
\(m_{\phi_W}=9,31\times10^{13}GeV\). The energy scale of Grand Unification
Theories
is \(m_{GUT}\sim 10^{14}-10^{16}GeV\) and the scale of
quantum gravity is \(m_{PL}
\sim 10^{19}GeV\). The strengths of the couplings between these spin-0
fields and
fermions are proportional to the masses of the fermions.
\item Due to the existences of three spin-0 states in the SM
new perturbation theory of the SM
is
constructed. The propagators of W and Z bosons are derived.
They have the same
expressions as the ones in renormalization gauge in the original
perturbation theory.
However, there are no additional ghosts. In the range of practical
energies Eqs.(72,100) are good
approximations.
\item The vacuum polarization by W and Z fields contribute to
renormalizations and
radiative corrections.
\item The effects of the spin-0 states can be found in loop diagrams
in which
propagators of intermediate bosons are involved. Studies of measurable effects
of spin-0 states are beyond the scope of this paper.
\item The interactions between $\phi^0$, $\phi^\pm$, fermions, and intermediate
bosons are shown in Eqs.(73,101,106).
\item In the new propagators of W and Z fields there are poles of
$\phi^{0,\pm}$ at very high energies $\sim10^{14}GeV$. On the other hand,
the minus signs of Eqs.(71,99) show that if these spin-0 states are on
mass shell there are problems of indefinite metric and negative probability
in the SM. These problems arise at energy scale of $10^{14}$GeV.
\end{enumerate}
\vspace{1cm}
The study is supported by a DOE grant.


\begin{thebibliography}{20}
\bibitem{} S.L.Glashow, Nucl. Phys. {\bf B22}, 579(1961);
S.Weiberg, Phys. Rev. Lett.
{\bf 19},1264(1967); A.Salam, Proc.of the $8^{th}$ Nobel Symposium, p.367,
ed.By
N.Svartholm, Almqvist and Wilsell, Stockholm, 1968.
\bibitem{} Bing An Li,
Proc. of the $29^{th}$ Intern. Conf. on High Nergy Physics, 23-29 July 1998, ed.
by
A.Astbyry, D.Axen, and J.Robinso, p.1603.
hep-ph/9903313; Nucl.Phys.{\bf B}(Proc. Suppl.)76,263(1999).
\end{thebibliography}
\end{document}